\documentclass[aip,cha,amsmath,amssymb,reprint]{revtex4-1}
\usepackage{graphicx}
\usepackage{dcolumn}% Align table columns on decimal point
\usepackage{bm}% bold math
\usepackage[colorlinks=true,linkcolor=blue]{hyperref}
\usepackage{times}
\expandafter\ifx\csname package@font\endcsname\relax\else
\expandafter\expandafter
\expandafter\usepackage
\expandafter\expandafter
\expandafter{\csname package@font\endcsname}%
\fi

\begin{document}

\title{An efficient, tunable, and robust source of narrow-band photon pairs at the $^{87}$Rb D1 line}
% Force line breaks with \\

\author{Roberto Mottola}
 \affiliation{Departement Physik, Universit\"at Basel, Klingelbergstr. 82, 4056 Basel, Switzerland.}
 \author{Gianni Buser}
 \affiliation{Departement Physik, Universit\"at Basel, Klingelbergstr. 82, 4056 Basel, Switzerland.}
 \author{Chris M\"uller}%
 \affiliation{ Institut f\"ur Physik, Humboldt-Universit\"at zu Berlin, Newtonstr. 15, 12489 Berlin, Germany.}
\author{Tim Kroh}%
\affiliation{ Institut f\"ur Physik, Humboldt-Universit\"at zu Berlin, Newtonstr. 15, 12489 Berlin, Germany.}
\author{Andreas Ahlrichs}%
\affiliation{ Institut f\"ur Physik, Humboldt-Universit\"at zu Berlin, Newtonstr. 15, 12489 Berlin, Germany.}
\author{Sven Ramelow}%
\affiliation{ Institut f\"ur Physik, Humboldt-Universit\"at zu Berlin, Newtonstr. 15, 12489 Berlin, Germany.}
\affiliation{IRIS Adlershof, Humboldt-Universit\"at zu Berlin, Zum Gro{\ss}en Windkanal 6, 12489 Berlin, Germany.}
\author{Oliver Benson}
\affiliation{ Institut f\"ur Physik, Humboldt-Universit\"at zu Berlin, Newtonstr. 15, 12489 Berlin, Germany.}
\affiliation{IRIS Adlershof, Humboldt-Universit\"at zu Berlin, Zum Gro{\ss}en Windkanal 6, 12489 Berlin, Germany.}
\author{Philipp Treutlein}
\affiliation{Departement Physik, Universit\"at Basel, Klingelbergstr. 82, 4056 Basel, Switzerland.}
\author{Janik Wolters}
\email{janik.wolters@dlr.de}
\affiliation{Departement Physik, Universit\"at Basel, Klingelbergstr. 82, 4056 Basel, Switzerland.}
\affiliation{Deutsches Zentrum f\"ur Luft- und Raumfahrt e.V. (DLR), Institute of Optical Sensor Systems,  Rutherfordstr. 2, 12489 Berlin, Germany.}

\date{\today}% It is always \today, today,

\begin{abstract}
We present an efficient and robust source of photons at the $^{87}$Rb D1-line (795 nm) with a narrow bandwidth of $\delta=226(1)$~MHz. The source is based on non-degenerate, cavity-enhanced spontaneous parametric down-conversion in a monolithic optical parametric oscillator far below threshold. The setup allows for efficient coupling to single mode fibers. A heralding efficiency of $\eta_{\mathrm{heralded}}=45(5)$~\% is achieved, and the uncorrected number of detected photon pairs is $3.8 \times 10^{3}/(\textrm{s mW})$. For pair generation rates up to $5\times 10^{5}/$s, the source emits heralded single photons with a normalized, heralded, second-order correlation function $g^{(2)}_{c}<0.01$. The source is intrinsically stable due to the monolithic configuration. Frequency drifts are on the order of $\delta/20$ per hour without active feedback on the emission frequency. We achieved fine-tuning of the source frequency within a range of $ > 2$~GHz by applying mechanical strain.
\end{abstract}

\maketitle
\onecolumngrid
\section{\label{sec:intro}Introduction}

Photonics is among the most promising platforms for realizing quantum technology, in particular quantum communication networks, quantum simulators and computers, and quantum sensors. Research is driven by manifold benefits: quantum repeater networks promise unconditional secure communication \cite{Sangouard2011, Kimble2008, Gisin2002}, quantum simulators and computers further better understanding of complex quantum systems, e.g.~in chemistry or many-body physics 
\cite{Buluta2009, Georgescu2014, Reiher2017, Cao2018, Ladd2010, Kok2007, OBrien2007}, 
and quantum sensors enable measurements of unprecedented sensitivity \cite{Aasi2013, Pirandola2018}.
To realize these applications, excellent single-photon sources are required. Ideally, such single-photon sources emit one photon at a time, on demand, at a high generation rate, and in a pure state of a single spatial, temporal, and spectral mode. Moreover, each copy of such a source should be capable of generating photons identical to those produced by the others \cite{Flamini2018}. Current sources can only fulfill a limited number of the above requirements while maintaining high performance for specific applications. Among the most advanced single photon sources are semiconductor quantum dots \cite{Senellart2017, Reigue2017, Thoma2016, Beguin2018, Lodahl2017} and sources based on spontaneous parametric down conversion (SPDC)\cite{Scholz2009, Ahlrichs2016, Fekete2013,Luo2015,Chuu2012}. 
A major issue of SPDC sources is that the photon generation probability per coherence time $\varepsilon$ has to be kept well below unity to avoid contamination with higher photon number states.
When $N$ photons are needed simultaneously to feed a quantum photonic circuit, the success probability scales exponentially as $P \sim \varepsilon ^N$. In typical state-of-the-art experiments, five-photon events occur at rates below $1/$s, \cite{Zhong2018} and ten-photon events are observed every few hours\cite{Chen2017}. This problem can be circumvented by massive multiplexing \cite{Migdall2002,Shapiro2007,Sinclair2014,Kaneda2015,Rambach2016,Joshi2018}, in which case the stated demands on ideal single photon sources apply to each mode individually, or by combining a heralded source with a memory that can store and release photons in a controlled way \cite{Jeffrey2004, Pittman2002, Nunn2013, Pang2018, Kaneda2017}.

Several quantum memory schemes have been developed and demonstrated for single photon storage\cite{Kaczmarek2018, Seri2017, Akiba2009}, and at least one proof of principle experiment \cite{Makino2016} has shown enhanced coincidence rates through synchronization in a read-only \cite{Afzelius2015} memory. However, even in the latter experiment the rates were far lower than what faster sources provide coincidentally without synchronization, compare \cite{Zhong2018}. A demonstration of practical synchronization of SPDC sources with the promise of scaling well beyond what is achievable without synchronization is still outstanding. For this, the memory storage time must be longer than the average time between two subsequently generated photons, i.e. the inverse of the photon pair rate, the heralding efficiency must be sufficient to overcome the memory readout noise, and it must in principle be possible to achieve a high efficiency and a large time-bandwidth product in the memory simultaneously. Moreover, for combining an SPDC source and memory to a functional, on demand, compound photon source, both the heralding efficiency and the memory efficiency should approach unity.

In our experiments we aim for an SPDC source that can be combined with broadband EIT-like quantum memories in warm atomic vapors (broadband with respect to the intrinsic linewidth of the atoms), as investigated by us \cite{Wolters2017} and others \cite{Saunders2016,Kupchak2015,Finkelstein2018}. The requirements imposed by the memory presented in Ref. \cite{Wolters2017} on the SPDC source are: 1. Tunable emission frequency near the D1 line of $^{87}$Rb. 2. Photon bandwidth between 100 MHz and 1 GHz. 3. Heralding efficiency above 25 \% such that the expected signal exceeds the memory readout noise. Additionally, we have set a goal of a heralding rate $> 10^{5}$ per second to be in range of practically working photon synchronization for a memory with 10 $\mu$s storage time, which was not demonstrated in \cite{Wolters2017}, but is known to be easily achievable in vapor cell memories. 

To efficiently generate photons with a bandwidth between 500 kHz and 1 GHz at tunable wavelength, resonator enhanced sources have been developed~\cite{Scholz2009, Ahlrichs2016, Fekete2013, Rambach2016}. At their core, these are optical parametric oscillators (OPO) pumped far below threshold \cite{Lu2000,Kuklewicz2006}. By increasing the electric field per photon, the triple resonant cavity enhances the generated pair rate per mW of pump power and simultaneously forces the emitted light into narrow spectral lines. These lines are arranged in clusters due to the different refractive index and free spectral range (FSR) for signal and idler polarizations \cite{Ahlrichs2016, Jeronimo-Moreno2010, Luo2015, Chuu2012}. The theory of such sources has been developed e.g. in Ref. \cite{Herzog2008, Jeronimo-Moreno2010}. Typical OPO photon pair sources are prone to intra-cavity losses and wavefront deformation that limited the coupling efficiency to single mode fibers to about 20 \%. 
To overcome these issues we follow a fully monolithic approach to build a triple resonant OPO. Double resonant monolithic OPOs \cite{Mehmet2010,Breitenbach1995,Brieussel2016,Schiller1996} and triple resonant semi-monolithic OPOs \cite{Mehmet2011, Guo2002, Zhou2015} have been investigated in the context of squeezing. Furthermore, the stability and tunability of photon pair generators through temperature and pump frequency has been studied in double resonant  \cite{Pomarico2009,Pomarico2012} and single resonant monolithic waveguide resonators  \cite{Ikuta2019}, respectively. A few experiments realized photon pair generation in a tunable, triple resonant, monolithic, disk-shaped resonator, but could not demonstrate the large heralding efficiencies and high pair generation rates required \cite{Fortsch2013,Schunk2016}. With our experiments we overcome these limitations. We present a monolithic, cavity-enhanced, non-degenerate SPDC source that generates signal photons at the $^{87}$Rb D1 line (795 nm), with a spectral bandwidth of 226(1)~MHz and a heralding efficiency of $\eta_{\mathrm{heralded}}=45(5)$~\%. The heralding rate approaches 
$10^{5}$ heralding events per second. The source can be fine-tuned over a range of $>2$~GHz by applying mechanical strain. The passive stability is limited by long-term drifts on the order of 10~MHz/h. As such, the presented source is directly compatible with the atomic vapor cell quantum memory from Ref. \cite{Wolters2017}. 
\section{\label{sec:setup}Experimental Setup}
\begin{figure}
\centering
\includegraphics[width=0.8 \columnwidth]{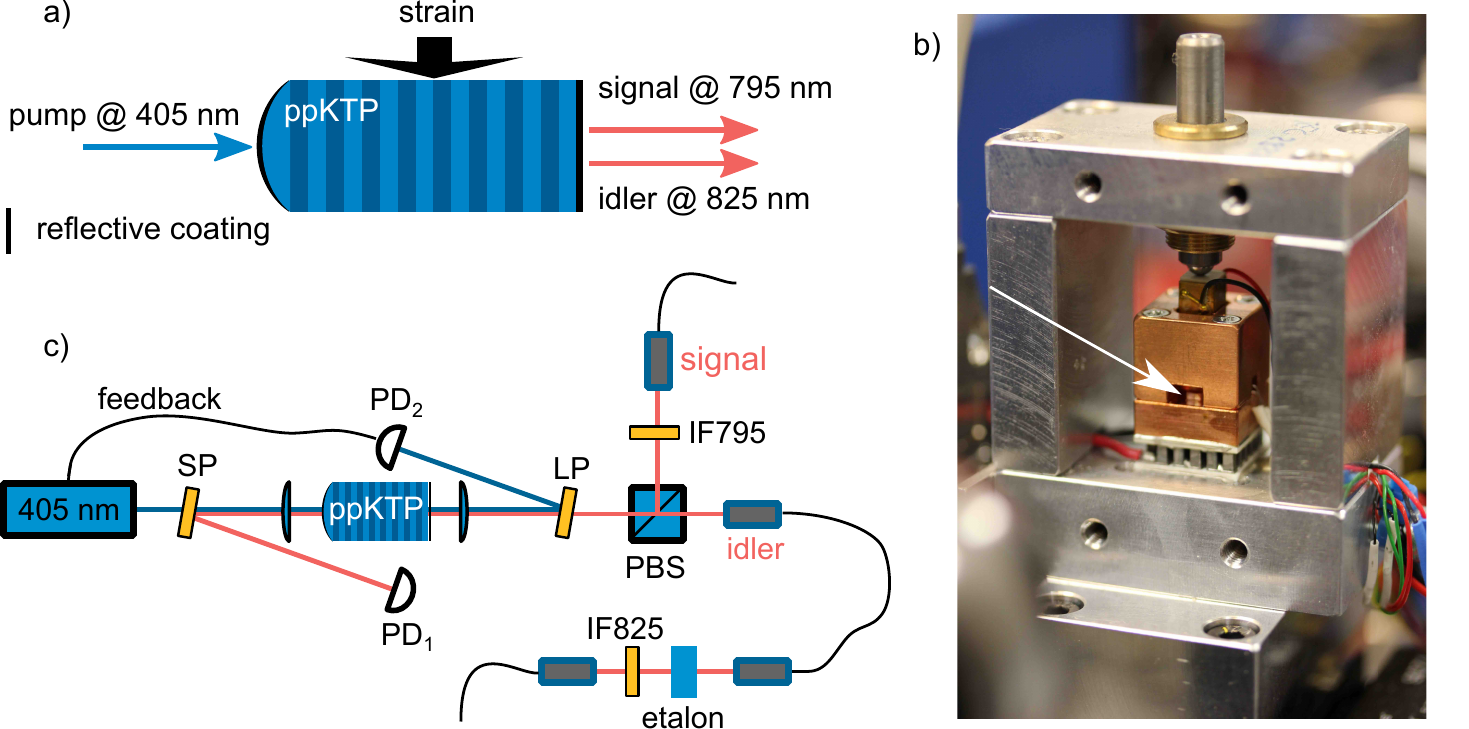}% Here is how to import EPS art
\caption{\label{fig:setup} The experimental setup. (a) Sketch of the monolithic OPO that is tuned by applying mechanical strain and controlling the temperature. Reflective coatings are applied on both end facets of the ppKTP crystal to form an optical cavity for the pump, signal, and idler wavelengths. (b) Photograph of the oven (housing partially removed) that contains the ppKTP crystal. Mechanical strain can be applied via a fine-threaded screw and piezo. The arrow indicates the position of the ppKTP crystal. (c) The optical setup used to pump the OPO and to collect signal and idler photons into optical fibers. SP, LP, IF: short-pass, long-pass, and band-pass interference filters respectively. PD: photodiode. PBS: polarizing beamsplitter.}
\end{figure}
The triple resonant OPO consists of a single periodically poled Potassium Titanyl (ppKTP) crystal (\textit{Raicol Crystals}), sized $1\times2\times7$~mm$^{3}$, with a 10.1~$\mu$m poling period for type-II conversion of 405~nm pump photons to 795~nm (825~nm) signal (idler) photons. The crystal forms a hemispherical Fabry-P\'erot resonator where the spherically convex facet has a 10~mm radius of curvature, see Fig.~\ref{fig:setup}(a). Dielectric coatings were applied with nominal reflectivities of $R_{s/i}^{c}=99.9$~\% for the signal/idler and $R_{p}^{c}=90$~\% for the pump wavelength on the convex facet, and $R_{s/i}^{p}=91.5$~\% and $R_{p}^{p}=99$~\% on the planar facet. Before coating the crystal, we measured an absorption coefficient of 1 dB per cm at 405~nm, and we assume vanishing absorption around 800 nm. Accounting for losses in the material, this allows for critical coupling of the pump light, while virtually all signal and idler photons leave through the planar surface. The crystal is sandwiched between two highly polished copper plates to apply mechanical strain perpendicular to the optical axis for fine-tuning of the triple resonance condition, see Fig.~\ref{fig:setup}(b). Using a thermoelectric element and a digital temperature controller, the mechanical mount is temperature controlled and stabilized to $\pm 5$~mK. For optical pumping of the OPO, a grating stabilized 405 nm diode laser provides up to 100 mW of optical power. 
Due to the finite reflectivity of the mirror on the planar ppKTP crystal facet some transmission of the pump laser occurs on resonance. The transmitted light is detected by an amplified photodiode (PD$_{2}$ in Fig.~\ref{fig:setup}(c)) and used to stabilize the pump laser frequency on the desired OPO resonance. No further active feedback is sent to the pump laser.
The signal and idler photons are separated on a polarizing beamsplitter and coupled into polarization maintaining single mode fibers with anti-reflection coated facets. Prior to the detection by avalanche photodiodes (APDs, Excelitas SPCM), the signal photons are filtered by a narrow-band (0.44~nm FWHM, manufacturer measurement) interference filter (IF) with $> 90$~\% transmission, while the idler photons are filtered by a temperature stabilized etalon (FSR~=~12.8(2)~GHz, calculated from 8.0(1)~mm specified thickness; FWHM~=~274(4)~MHz, measured using the FSR as frequency scale) and an IF with 0.57(5)~nm FWHM measured at 822~nm. The filtering stage in the idler arm reaches a peak transmission of $\approx 80$~\%. Its extinction is the product of the extinction of the IF, which is $<10^{-2}$ for frequencies detuned by at least one linewidth from the center, and the extinction of the etalon, which is $1.2\times 10^{-3}$ at the FSR$/2$ frequencies. As the bandwidth of this etalon is comparable to the linewidth of the source it has an impact on the detection rate of the idler photons, transmitting about 68~\% of the photons in a 400~MHz window around the central frequency. 
In future experiments on the storage of single photons the quantum memory will act as an additional spectral filter in the signal arm, as light away from resonance will not interact with the atoms. Moreover, all memory experiments involve some kind of spectral filtering by etalons that will remove any residual signal photons that are detuned from the desired signal frequency. 
A tunable 795~nm diode laser that is referenced to the $^{87}$Rb D1 line can be injected into the OPO via the signal arm to measure the transmission spectrum on PD$_{1}$ for adjustment and characterization. The same laser is simultaneously used to stimulate difference frequency generation (DFG) at the idler wavelength as a tool for measuring the spectrum of the OPO. For detecting the DFG signal, an amplified photodiode is inserted into the idler arm prior to the filtering stage.

\section{\label{sec:results}Results}

\begin{figure}
\centering
\includegraphics[width=0.8 \columnwidth]{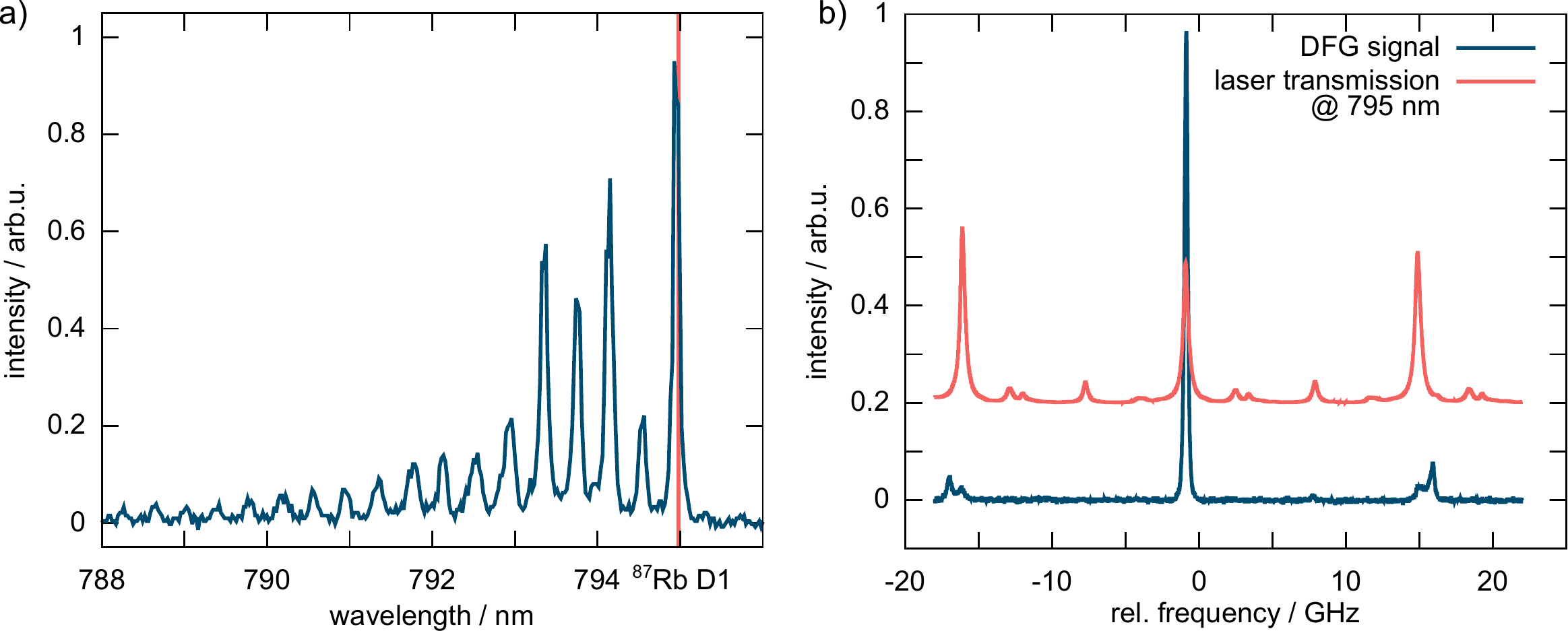}\\
\caption{\label{fig:spectrum}The spectral properties of the source. (a) Spectrum of the signal photons after coarse tuning, recorded on a 500 mm spectrometer with a CCD camera sensitive to single photons. The vertical red line indicates the position of the $^{87}$Rb D1 line at 794.979 nm. The measurement was taken at a pumping power of approximately 1 mW. (b) Transmission spectrum of the ppKTP cavity around the $^{87}$Rb D1 line, recorded on PD$_1$ and the corresponding DFG intensity measured simultaneously on an amplified photodiode in the idler arm. In the 795 nm laser transmission, three major peaks corresponding to TEM$_{00}$ modes of the cavity are visible. The free spectral range (FSR) is 16~GHz. The DFG signal shows the fine structure of one emission cluster and consists of a central peak accompanied by two weaker doublets each separated by one FSR. Outside the displayed frequency window no DFG signal was measurable.}
\end{figure}

For coarse adjustment of the signal frequency the IF is removed and a spectrum of the signal photons is recorded using a 500~mm grating spectrometer with a cooled CCD camera. The signal spectrum (Fig.~\ref{fig:spectrum}(a)) spans across the phase matching bandwidth of a few nm and shows several sharp lines separated by 0.5~nm. Each line corresponds to one emission cluster. The fine structure of the individual clusters is not resolved. 
The wavelength of the brightest cluster at the red side of the spectrum is tuned to be in close proximity to the $^{87}$Rb D1 line, which is always possible by adjusting the crystal temperature and pumping wavelength.
Subsequently, the OPO is seeded by the tunable 795~nm laser. This allows us to measure the OPO transmission, as well as to record the intensity of the DFG signal while the tunable laser is scanned by $\pm 20$~GHz around the $^{87}$Rb D1 transitions, see Fig.~\ref{fig:spectrum}(b). This resolves the fine structure of the emission cluster. 
Due to the large difference in the FSR for signal and idler polarizations, the cluster consists of one strong central peak for which the signal and idler resonances are well aligned and two weaker doublet peaks. Since the pump is locked to the cavity resonance, triple resonance is achieved at the central mode. 
The central peak has a bandwidth of $\delta= 226(1)$~MHz (FWHM), determined by the ppKTP cavity linewidths for signal and idler. Due to the comparably large difference in FSR for signal and idler, the central peak contributes 75~\% to the overall intensity of the cluster. 
The filtering stage in the idler arm is adjusted to transmit the central DFG peak. Subsequently, the seeding laser is switched off and single photons in the signal and idler arms are measured by time-tagged time-resolved (TTTR) photon counting. At a pump power of $P_{\mathrm{pump}}= 1.2(1)$ mW, we typically detect photons in the signal and idler arm at a rate of $n_{s} = 6.8 \times 10^{4}~\textrm{counts}/\textrm{s}$ and $n_{i} = 1.7\times 10^{4}~\textrm{counts}/\textrm{s}$, respectively. The measured idler rate is much lower due to more stringent spectral filtering to one single emission line. Integrating the total number of coincidences around the detection time of the idler photon, we obtain a detected signal-idler pair rate of $r = 4.6\times 10^{3}~\textrm{pairs}/\textrm{s}$, or $3.8 \times 10^{3}/(\textrm{s mW})$ when normalized to pump power. From this we find $\eta_{s} = \frac{r}{n_{i}}= 27$~\% and $\eta_{i} = \frac{r}{n_{s}}=6.7$~\% for the efficiencies in the signal and idler arm respectively. When correcting for the detector efficiency of $\eta_{\mathrm{det}} = 60(6)$~\%, we estimate the heralding efficiency for the signal arm to be $\eta_{\mathrm{heralded}}= \frac{r}{\eta_{\mathrm{det}} n_{i}}=45(5)$~\%. Up to $\eta_{\mathrm{heralded}}=50(5)$~\% has been observed, i.e.~detection of an idler photon indicates the presence of a signal photon in the polarization maintaining single mode fiber with a probability of about 1/2.
The total rate at which photon pairs are generated inside the cavity, normalized to $P_{\mathrm{pump}}$, is $R=r/(P_{\mathrm{pump}}\eta_{s}\eta_{i})=n_{i}n_{s}/(r P_{\mathrm{pump}})=2.1(2)\times 10^{5}~\textrm{pairs}/(\textrm{s mW})$, where the uncertainty is due to the measurement uncertainty of the pump power.

\begin{figure}
\centering
\includegraphics[width=0.8\columnwidth]{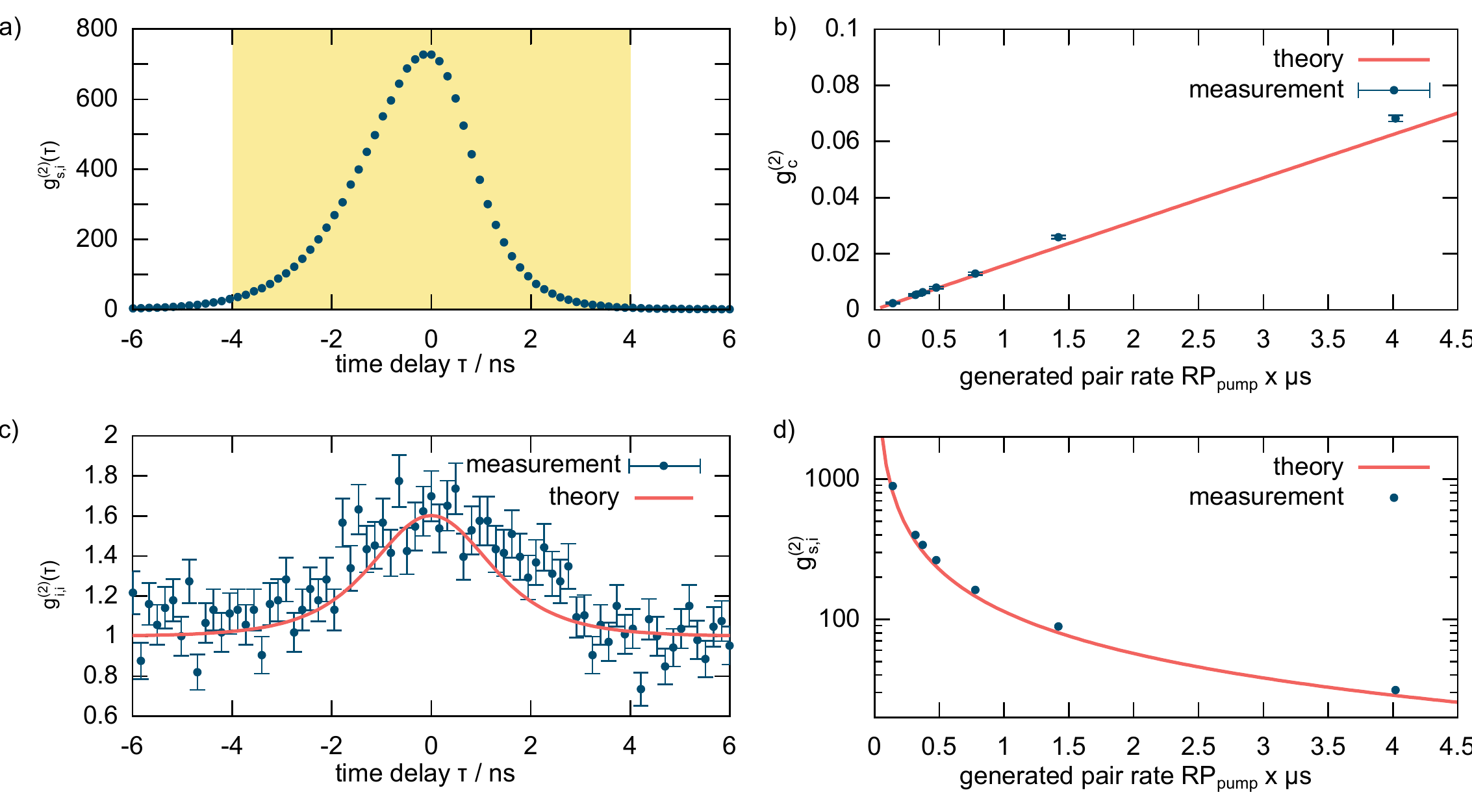}\\
\caption{\label{fig:crosscorr} Second-order correlations of the generated photons. (a) Measured second-order cross-correlation between signal and idler photons $g^{(2)}_{s,i}(\tau)$ ($162$~ps bins). The pronounced bunching at zero time delay indicates that photons are emitted in pairs. The yellow shaded area shows the coincidence window with $\Delta t = 98 \times 81$~ps~$\approx8$~ns used for evaluating the second-order correlation shown in (b). The measured shape follows a double exponential, convoluted with the instrument response function. The temporal asymmetry reflects the dissimilar frequency filtering in signal and idler arms. (b) Conditional second-order autocorrelation $g^{(2)}_{c}$ for various pair generation rates. Larger rates, practically equivalent to stronger pumping, come with an increased probability of multiphoton events. The source emits single photons with $g^{(2)}_{c}<0.01$ for generation rates up to $5\times 10^5$~pairs$/$s, i.e. up to 2.4~mW of pump power. (c) Unconditioned idler-idler autocorrelation $g^{(2)}_{i,i}(\tau)$ ($162$~ps bins). The measured data are in good accordance with the model (description in the text) taking the detector response function and the linewidth of the idler photons after filtering into account. (d) Signal-idler cross-correlation $g^{(2)}_{s,i}$ for different photon pair generation rates. The data are evaluated in $8$~ns bins. The theory and relation to b) is described in the text.}
\end{figure}

In order to characterize the quality of the photon pairs emitted by the source we measure the second-order correlations. The signal-idler cross-correlation $g^{(2)}_{s,i}(0)\gg2$ shows pronounced super-thermal bunching for zero time delay, see Fig.~\ref{fig:crosscorr}(a). This is a first indication of the generation of non-classical light. To estimate the single photon purity of the heralded single photons, i.e.~the degree of contamination with higher photon number states, we measure the time integrated second-order autocorrelation $g^{(2)}_{c}$ conditioned on the detection of an idler photon. To this end, we perform TTTR photon counting with three detectors: two (APD$_1$ and APD$_2$) in Hanbury Brown and Twiss configuration in the signal arm, and one (APD$_3$) in the idler arm. From the measured detection rates in the TTTR-data we estimate the single photon probability $P_{s_1}$ ($P_{s_2}$) for detection of one signal photon on APD$_1$ (APD$_2$) within a coincidence window of $\Delta t \approx 8$~ns upon the detection of an idler photon, as shown in Fig.~\ref{fig:crosscorr}(a), as well as the two photon probability $P_{d}$ to detect two signal photons upon the detection of an idler photon. From this, the measured conditional second-order autocorrelation $g^{(2)}_{c}= P_{d}/(P_{s_1}P_{s_2})$ is computed. 
The conditional second-order autocorrelation $g^{(2)}_{c}$ corresponds to the time average of the time dependent normalized Glauber autocorrelation function \cite{Fortsch2013} across the coincidence window. It gives the average single photon purity of a heralded photon.
We model the data with equation 24 derived in \cite{Sekatski2012}, for non-number resolving detectors. This theory is expressed in terms of the photon pair generation probability, which for CW pumping and in the regime far below threshold can be estimated as $p\approx RP_{\mathrm{pump}}\Delta t$ \cite{Herzog2008, Nielsen2007}. In the calculation of $R$ the inefficiencies cancel, and as the lowest count rates shown here are a factor 100 above the dark count rates, we examine the model in the limit of full efficiency and no dark counts. This yields the expression 
\begin{equation}
g^{(2)}_{c}=2RP_{\mathrm{pump}}\Delta t-(RP_{\mathrm{pump}}\Delta t)^2.
\end{equation}
Figure~\ref{fig:crosscorr}(b) shows that the measured $g^{(2)}_{c}$ is in agreement with the theoretical prediction.
For generation rates of up to $5 \times 10^{5}$~pairs$/$s, the source emits single photons with a $g^{(2)}_{c}<0.01$, proving that multi-photon generation is strongly suppressed. 

For light generated by SPDC the unconditioned second-order autocorrelation is an excellent measure for the number of modes $N$ emitted by the source, following the relation $g^{(2)}_{x,x}(0)=1 +1/N$ in the limit far below threshold \cite{Christ2011}. A careful accounting of mode number in an SPDC source can be found for instance in \cite{Rielaender2016}. However, due to timing jitter on the detectors we do not expect to measure $g^{(2)}_{i,i}(0)=2$ even for a single idler mode. 
The linewidth of the idler photons is determined by the product of the 226(1)~MHz Lorentzian line of the source cavity and the 274(4)~MHz Lorentzian line of the idler filter cavity. The relation between linewidth and coherence time for Lorentzian light is $\tau_0=(\pi\Gamma)^{-1}$ \cite{SalehTeich1991}, here $\tau_0=2$~ns. The ideal shape of the idler-idler autocorrelation is then $g^{(2)}_{i,i}(\tau)=1+\exp(-2\vert\tau\vert/\tau_0)$ \cite{GerryKnight2004}. We measure the combined timing jitter of our HBT configured detectors to be 1.72(4)~ns by measuring the width of the detectors' cross-correlation when 12~ps long, periodic pulses of 820~nm light are incident. The model plotted in Fig.~\ref{fig:crosscorr}(c) is a convolution of a Gaussian instrument response function accounting for this jitter and the ideal case. The agreement between the data and the model, as well as our characterization of the extinction ratio of the filters, leads us to believe that the remaining contamination of undesired modes in the idler is negligible in comparison to the detector induced uncertainty on the heralding efficiency. In contrast, as the signal arm is not significantly filtered, the additional modes wipe out the features in the unconditioned signal-signal autocorrelation, and $g^{(2)}_{s,s}(\tau)\approx 1$ everywhere, within measurement accuracy. When conditioning on the detection of an idler, the signal photons detected as coincidences inherit this enhanced mode purity, while the signal photons in other modes will register only as uncorrelated noise. From Bayes's Theorem the following relation between second-order correlation functions can be derived \cite{Chou2004}
\begin{equation}
g^{(2)}_{c} = \frac{g^{(2)}_{\vphantom{i} s,s} g^{(2)}_{i,i}}{g^{(2)}_{s,i}}.
\end{equation}
For the theory curve in Fig.~\ref{fig:crosscorr}(d) we insert Eq. 1 into Eq. 2 and substitute the idler autocorrelation for the signal to correct for the contamination by unfiltered modes. We again consider an 8~ns integration window, wherein we measure $g^{(2)}_{i,i}=1.338(16)$.

\begin{figure}
\centering
\includegraphics[width=0.8 \columnwidth]{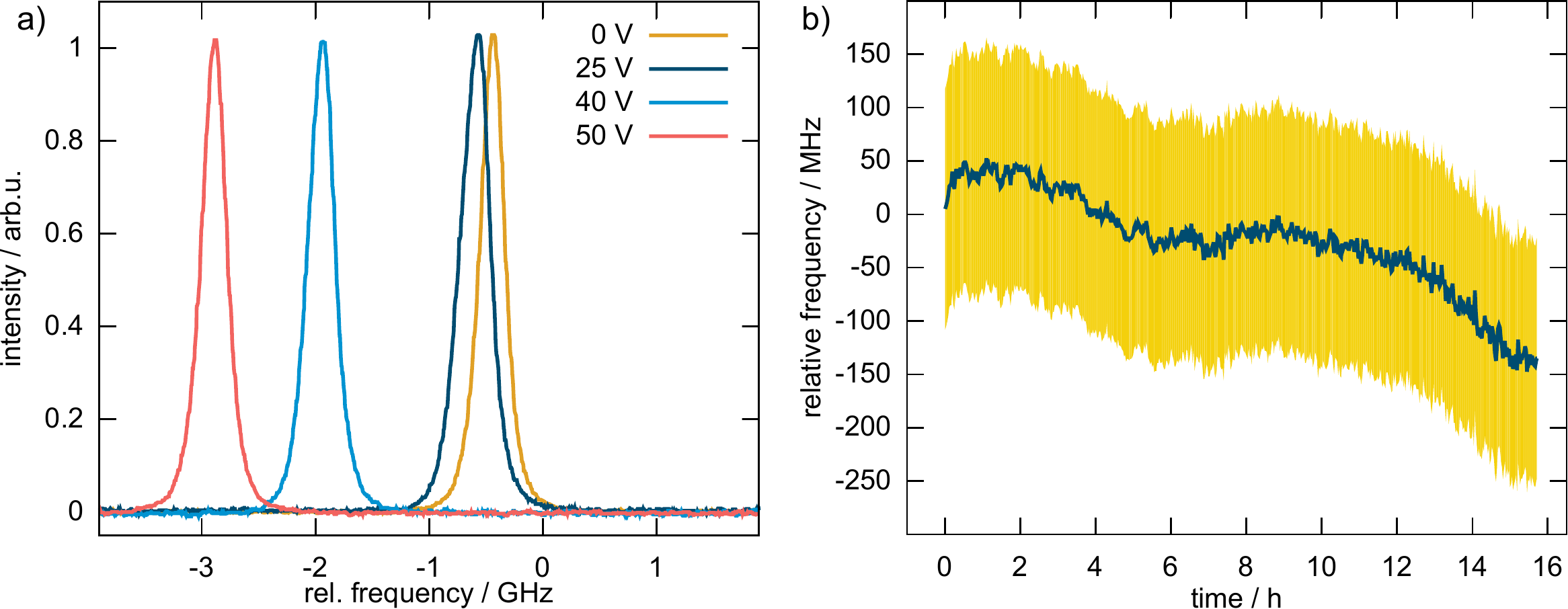}\\
\caption{\label{fig:tuning}Tuning behaviour and stability of the source. (a) Fine-tuning properties of the source. The measured DFG intensity depending on the seeding laser frequency relative to the $^{87}$Rb D1 line is shown for various piezo voltages. At $\approx$~20 V the piezo makes contact with the copper stamp and nearly linear tuning of the central emission peak by several GHz can be achieved.
	(b) Measurement of the source stability. While the source temperature is kept constant and the pump laser is stabilized to be on resonance with the OPO, the seeding frequency for maximum DFG signal is measured relative to the initial value (blue curve). The yellow shaded area indicates the FWHM linewidth of the DFG signal. The frequency measurement error is on the order of 5 MHz (not shown).}
\end{figure}

To fine-tune the emission frequency of the source, mechanical strain is applied onto the ppKTP crystal along the vertical axis\cite{Zielinska2018, Zielinska2017}. The strain deforms the refractive index ellipsoid and allows for fine-tuning of the triple resonance condition. To this end, a piezoelectric actuator presses a highly polished copper stamp onto the ppKTP crystal that is lying on a second polished copper surface, see Fig.~\ref{fig:setup}(b). By frequency tuning the seeding laser, changing the piezo voltage, and subsequently readjusting the crystal temperature by a few tens of mK triple resonance is reestablished. This temperature change also affects the phase matching condition. However, due to the phase matching bandwidth of several hundreds of GHz, this has only a minor effect and we achieved tuning by $> 2$~GHz from the initial value, as illustrated in Fig.~\ref{fig:tuning}(a).
After careful pre-adjustment this is sufficient to reach a desired frequency, e.g.~slightly red-detuned from the $F=1 \rightarrow F'=1$ transition of atomic $^{87}$Rb, where EIT quantum memories work preferentially \cite{Wolters2017, Namazi2017}.

To evaluate the frequency stability of the OPO, the source was operated continuously for 16 h while seeded by the tunable 795~nm laser. The laser frequency relative to the $^{87}$Rb D1 line was repeatedly scanned, and the frequency corresponding to the maximum DFG signal was recorded, see Fig.~\ref{fig:tuning}(b). We measured an average frequency drift of about 10 MHz per hour with no active feedback to control the emission frequency.

It remained possible to reach the $F=1 \rightarrow F'=1$ transition by strain and temperature tuning over a period of several weeks. However, on longer timescales changes of the triple resonance frequencies that could not be compensated by changing the piezo voltage and temperature occurred. We suspect that these changes are due to pump-induced gray tracking\cite{Boulanger1994,Jensen2007} that affects the cavity modes. Further investigations on this are beyond the scope of this letter. When the tuning range is thus exhausted a major readjustment is required to bring the source back within range of the desired atomic transition. To this end the pump frequency is changed by a few 100 GHz and the crystal temperature is adjusted accordingly. After readjustment the source performance was comparable to the described performance. Alternatively, the source can be pumped off-resonantly, however this requires larger pumping powers to reach the same pair rate.

We would like to point out that we investigated a second OPO formed by a 5 mm long, but otherwise identically specified, crystal from the same batch. With this second crystal we could achieve similar results, in particular tuning to the Rb D1 line was also possible.

A cavity enhanced narrow-band photon pair source based on a monolithic OPO was presented. The monolithic OPO can be tuned over several GHz, exhibits high stability, and, unlike preceding designs, does not need active stabilization of the OPO cavity. Moreover, intra-cavity surfaces that tend to introduce losses are avoided and consequently a high heralding efficiency of $\eta_{\mathrm{heralded}}=45(5)$~\% is reached. Given the relatively low finesse of $F = 36$ for the signal photons we expect the efficiency to be limited mainly by the fiber coupling.
In future experiments we will improve the surface quality of the ppKTP crystal, the dielectric coatings, and the collection optics, potentially comprising adaptive elements \cite{Minozzi2013}, to push the efficiency towards unity.
Even at comparably high photon pair generation rates of up to $R=5\times 10^{5}/$s the signal photons show pronounced single photon characteristics with measured $g^{(2)}_{c}<0.01$. 

The presented triple resonant monolithic OPO design can be adapted to provide photons for various experiments, e.g. on broadband light storage \cite{Saunders2016, Kupchak2015} or on combining dissimilar quantum systems. The presented source is directly compatible with the quantum memory from Ref. \cite{Wolters2017}, which will be used in future experiments to synchronize the probabilistic photon generation, e.g. for increasing the event rate in multi-photon experiments. This may lead to the development of nearly ideal compound photon sources that emit one photon at a time, on demand, and in a single spatial, temporal, and spectral mode, and that can be constructed in a reproducible fashion. Such a source would have high potential to enable photonic quantum simulation and computing with many photons.

\section*{\label{sec:funding}Funding}
Marie Sk\l{}odowska-Curie Actions of the EU Horizon 2020 Framework Programme (702304); 
Swiss National Science Foundation (SNF) (NCCR QSIT);
Emmy-Noether-Programm of the German Research Foundation (DFG) (RA 2842/1-1);
German Research Foundation (DFG) (CRC 787 project C2);
German Federal Ministry of Education and Research (BMBF) (Q.Link-X)

\section*{\label{sec:acknowledgments}Acknowledgments}
We thank Alisa Javadi and Natasha Tomm for use of their ps laser and help in determining the detector jitter, and Pavel Sekatski for helpful discussions on the theoretical models. 

\section*{\label{sec:disclosures}Disclosures}
The authors declare no conflicts of interest.

\bibliography{SPDC}

\end{document}